\documentclass[11pt]{article}

\usepackage{amssymb}
\usepackage{amsfonts}
\usepackage{amsmath}
\usepackage{graphicx}
\usepackage{latexsym}

\usepackage{hyperref}

\newcommand{\bb}{\begin{eqnarray}}
\newcommand{\ee}{\end{eqnarray}}
\newcommand{\beq}{\begin{equation}}
\newcommand{\eeq}{\end{equation}}
\newcommand{\ba}{\begin{array}}
\newcommand{\ea}{\end{array}}

\title{Darboux polynomial matrices: the classical Massive Thirring Model as study case}

\author{Antonio Degasperis\\
Dipartimento di Fisica, Universit\`a di Roma ``La Sapienza'', \\
E-mail: antonio.degasperis@roma1.infn.it}

\date{25-11-2014}

\begin{document}


\maketitle

\begin{abstract}
 One way of constructing explicit expressions of solutions of integrable systems of Partial Differential Equations (PDEs) goes via the Darboux method. This requires the construction of Darboux matrices. Here we introduce a novel algorithm to obtain such matrices in polynomial form. Our method is illustrated by applying it to the classical Massive Thirring Model (MTM), and by combining it with the Dihedral group of  symmetries possessed by this model.

\vspace{0.2cm}

\end{abstract}


\section{Introduction}
Several systems of coupled nonlinear Partial Differential Equations (PDEs) have been proved to be both integrable and good (though approximate) models of a variety of wave propagation phenomena. The mathematical property of integrability, together with the broad range of physical contexts in which integrable models find their application, are the main ingredients of  \emph{soliton theory}. This subject is almost half-century old and has motivated such a vast production of results that we limit ourselves to quote only those references which are strictly related to our present work. In soliton theory a particularly fruitful method of constructing special wave solutions (mainly soliton solutions) is that introduced by Darboux in 1882 while investigating Ordinary Differential Equations (ODEs). If applied to the Lax pair of matrix ODEs, this method yields explicit expressions of parametric families of soliton solutions (see f.i. \cite{MS}) in both the isospectral evolution case (which is the one considered here) and non isospectral  evolution case. The basic point of this method is the covariance (form-invariance) of a linear homogeneous matrix ODE of the form
\begin{equation}\label{ode}
\Psi_\xi = A(\zeta,\xi) \Psi
\end{equation}  
with respect to a linear transformation (Darboux Transformation, DT) of the dependent variable $\Psi\rightarrow \Psi' = D(\zeta,\xi) \Psi$. Thus the covariance requirement is
\begin{equation}\label{newode}
\Psi'_\xi = A'(\zeta,\xi) \Psi' \;\;.
\end{equation} 
with additional conditions as specified here below. 
The real variable  $\xi$ is the independent one, the subscript standing for differentiation, the coefficient $A(\zeta,\xi)$ as well as the transformed coefficient $A'(\zeta,\xi)$ (and therefore also $\Psi$,  $\Psi'$, and the DT matrix $D(\zeta,\xi)$) are $M \times M$ matrices. Moreover the matrix $A(\zeta,\xi)$ is assumed to have a rational dependence on the (so-called spectral) complex parameter $\zeta$, namely
\begin{equation}\label{rational}
 A(\zeta,\xi) = \sum_{j=0}^n A_j (\xi) \zeta^j + \sum_{k=1}^l \frac{Q_k(\xi)}{\zeta-\alpha_k} 
\end{equation}
and the Darboux matrix $D(\zeta,\xi)$ is required to maintain this property, namely
\begin{equation}\label{newrational}
 A'(\zeta,\xi) = \sum_{j=0}^n A'_j (\xi) \zeta^j + \sum_{k=1}^l \frac{Q'_k(\xi)}{\zeta-\alpha_k} \;\;.
\end{equation}
Finally, and most crucial, the Darboux matrix itself is asked to have a rational dependence on $\zeta$,
 \begin{equation}\label{Drational}
 D(\zeta,\xi) = \sum_{j=0}^N D^{(j)} (\xi) \zeta^j + \sum_{k=1}^L \frac{R_k(\xi)}{\zeta-\beta_k} \;\;.
\end{equation}
As a matter of fact, this rational structure of the Darboux matrix can be changed, without affecting the action of the DT itself, by multiplying the matrix $D(\zeta,\xi)$ by an arbitrary scalar $\xi-$independent  function $f(\zeta)$. This way the matrix (\ref {Drational}), according to personal taste, may be given  the polynomial form
 \begin{equation}\label{Dpoly}
 D(\zeta,\xi) = \sum_{j=0}^{N+L} D^{(j)} (\xi) \zeta^j  \;\;,
\end{equation}
or the simple-pole expression
 \begin{equation}\label{Dpole}
 D(\zeta,\xi) = D_0(\xi) + \sum_{k=1}^{N+L} \frac{R_k(\xi)}{\zeta-\beta_k} \;\;.
\end{equation}
or any intermediate form like (\ref {Drational}). While dealing with concrete cases, computational tasks aimed to construct $D(\zeta,\xi)$  may suggest the form of this matrix which seems to be more convenient. By and large, the pole expansion (\ref{Dpole}) is naturally close to the dressing method \cite{ZMNP}, or to the inverse spectral technique (see f.i. \cite{CD}), and, for an instructive review of the Darboux method in this connection,  see also \cite{C2009}.  As for the polynomial form (\ref{Dpoly}), its use is perhaps less common and its formulation for an arbitrary degree of $D(\zeta,\xi)$ can be found for instance in \cite{NM,SMN} (see also \cite{MS,C2009}). 

\noindent
The purpose of the following section is introducing the Lagrange form of the Darboux matrix which is alternative to the power expansion representation (\ref{Dpoly}). In fact the choice of the representation is a general issue concerning any matrix valued polynomial and has no relation with the Lax pair of ODEs. Combining this representation with the Lax pair is the main content of the next section 3. Our treatment there will be focused on the (classical) Massive Thirring Model (MTM) and confined, as an illustration of our method, to the simplest DT associated with the Lax pair. Our interest in this integrable model is not only motivated by its field theoretical role as a nonlinear Dirac spinor system (which is the relativistic counterpart of the Nonlinear Schr\"{o}dinger equation), but also by the fact that it may be considered close enough to propagation equations of two coupled optical modes in nonlinear periodic media \cite{AW,MAKKC}. The construction of the $N-$soliton solution, over both the vacuum and the continuous wave background, have been done by the dressing method (say by the simple-pole formula (\ref{Dpole}) of the Darboux matrix). Thus the expression of these solutions is already available \cite{DHS,BGK}, and we limit ourselves to constructing the lowest degree polynomial Darboux matrix to point out the novel features of our technique.  While doing this we also combine our treatment with the symmetries of the MTM which are related to the dihedral symmetry group  \cal{D}$_2$ of the corresponding Lax pair (for the dihedral group as reduction group see \cite{LM,MPW}).    


\section{ Lagrange form of polynomial matrices}
To the purpose of  introducing a new method of constructing polynomial Darboux matrices,  we preliminarily  consider in this section the Lagrange interpolation algorithm as applied to matrix valued polynomials.  Let $\{\zeta_1\,,\,\zeta_2\,,\,\cdots\,,\,\zeta_N\}$ be a given set of $N$ complex numbers such that $\zeta_j \neq \zeta_k$ if $j\neq k$. In the space of $(N-1)-$degree polynomials of $\zeta$ the Lagrange basis is defined as
\begin{equation}\label{lagrange}
L^{(N)}_j(\zeta) = \frac{Q^{(N)}_j(\zeta)}{Q^{(N)}_j(\zeta_j)} \;,\; j=1,\dots,N \;\;,\;\;
Q^{(N)}_j(\zeta)= \prod_{k\neq j}^N (\zeta-\zeta_k)\;\;.
\end{equation}
The $N$ Lagrange polynomials $L^{(N)}_j(\zeta)$ are such that 
 \begin{equation}\label{basis}
L^{(N)}_j(\zeta_k) = \delta_{jk}\;\;,\;\; j\,,\,k\,=1,\dots,N \;\;,
\end{equation}
where the right hand side is the Kronecker delta. We note here that the $N$ coefficients $C^{(N)}_{jk}$ of $L^{(N)}_j(\zeta)$,
\begin{equation}\label{lagrangecoeff}
L^{(N)}_j(\zeta) =  \sum_{k=1}^N C^{(N)}_{jk} \zeta^{k-1}\;\;,
\end{equation}
are the matrix elements  of the inverse $C^{(N)}$ of the $N\times N$ Vandermonde matrix $V^{(N)}(\zeta_1,\cdots \zeta_N)$, whose entries are $V^{(N)}_{jk}=\zeta_j^{k-1}$, namely $C^{(N)}=V^{(N)-1}$. In fact,  relations between Darboux matrices and Vandermonde, or Vandermonde-like matrices, have been already pointed out \cite{SMN}. Here our main use of the Lagrange polynomials (\ref{lagrangecoeff}) is the expression
\begin{equation}\label{lagrangeinter}
D(\zeta)= \sum_{j=1}^N L^{(N)}_j(\zeta) D_j  \;\;,
\end{equation}
of the $M \times M$ matrix valued polynomial $D(\zeta)$ of degree $N-1$ whose values $D_j$ at $N$ given points $\zeta_j$ of the complex plane, $D(\zeta_j)=D_j$, are given. 
For future reference, we observe that if the polynomial $D(\zeta)$ is of degree $S$ (as we assume from now on), with $S\,<\,N-1$, then the expression (\ref{lagrangeinter}) still holds provided the given matrices $D_j\,,\, j=1,\dots,N$ satisfy the $N-S-1$ conditions
\begin{equation}\label{matrixcond}
 \sum_{j=1}^N C^{(N)}_{jk+1} D_j =0 \;\;,\;\; k=S+1\,,\,\dots\,,\,N-1\;\;.
\end{equation} 
It is obvious that these conditions are identically satisfied if the matrix $D_j$ were set a priori as the value of the given polynomial $D(\zeta)$ at $\zeta = \zeta_j$. Instead, if   
 $D_j$ is chosen by a different criterium, as we do below, these conditions (\ref{matrixcond}) become significant and crucial to our method. 
This observation is relevant to the next step that is assigning the $N$ points $\zeta_j$ as the roots, which are assumed to be \emph{simple}, of the polynomial
 \begin{equation}\label{roots}
P(\zeta) = \textrm{det}[D(\zeta)] \;\;,\;\; P(\zeta_j)= 0   \;\;.
\end{equation} 
Thus $N=MS$ and the expression (\ref{lagrangeinter}) is the $(MS-1)-$degree Lagrange form of the $S-$degree polynomial $D(\zeta)$. This implies that the given matrices $D_j$ have to satisfy the $S(M-1)-1$ conditions (\ref{matrixcond}) with $N=MS$. Since, by construction, det$D_j =0$, for each $j$ there exists a $M-$dimensional non vanishing vector $z_j$ such that 
 \begin{equation}\label{vecv}
D_j z_j=0  \;\;,\; j=1,\dots,MS \;\;.
\end{equation} 
 We further assume that $z_j$ is \emph{simple}, namely that ker$[D_j]$ is a one dimensional subspace. Associated with each vector $z_j$ we introduce a basis of $M-1$ vectors, $\{y^{(1)}_j\,,\cdots,\,y^{(M-1)}_j \}$ of the subspace which is orthogonal to $z_j$, say
 \begin{equation}\label{orthog}
y^{(\alpha)\dag}_j z_j= 0   \;\;,\;\;\alpha = 1,\dots,M-1\;\;.
\end{equation}   
Hereafter any complex vector $v$ is treated as a one-column matrix and therefore its Hermitian conjugate $v^{\dag}$ is a one-row matrix. The standard scalar product of two vectors $y$ and $v$ is $y^{\dag} v$ while the dyadic product $v y^{\dag} $ is a square matrix. With this notation, any matrix $D_j$ can be parametrized as
 \begin{equation}\label{Dj}
D_j=\sum_{\alpha=1}^{M-1} w^{(\alpha)}_j y^{(\alpha)\dag}_j     \;\;,\; j=1,\dots,MS \;\;,
\end{equation}   
since this expression of $D_j$ satisfies (\ref{vecv}) for any choice of the $M-1$ linearly independent arbitrary vectors $w^{(1)}_j\,,\cdots,\,w^{(M-1)}_j$. Note that, if, for any $j$, the vectors $z_j$ and $y^{(1)}_j\,,\cdots,\,y^{(M-1)}_j$ are fixed, then the matrix $D_j$ is parametrized, see (\ref{Dj}), by  $M(M-1)$ complex numbers, namely the components of the $M-1\; M-$dimensional vectors $w^{(\alpha)}_j$. 

\noindent
At this point it is worth noticing that we have obtained a particular Lagrange representation (\ref{lagrangeinter}) of a $S-$degree polynomial matrix $D(\zeta)$ by choosing the $N=MS$ points $\zeta_j$ according to the prescription (\ref{roots}), and by consequently expressing the values $D(\zeta_j)=D_j$ as given by (\ref{Dj}). On the other hand the polynomial $D(\zeta)$ has the standard power representation
\begin{equation}\label{poly}
D(\zeta)= \sum_{k=0}^S D^{(k)} \zeta^k  \;\;,
\end{equation}
whose $S+1$ coefficients $D^{(k)}$ are $M \times M$ matrices. These two equivalent representations of the same polynomial matrix $D(\zeta)$ imply a  relation between the matrices $D_j$, see (\ref{Dj}), and the coefficients $D^{(k)}$, see (\ref{poly}). This relation follows from (\ref{lagrangecoeff}), (\ref{lagrangeinter}) and (\ref{poly}) and reads
\begin{equation}\label{matrixrel}
D^{(k)} =  \sum_{j=1}^{MS} C^{(MS)}_{jk+1} D_j\;\;,\;\; k=0,\dots,S\;\;.
\end{equation}
In addition the matrices $D_j$ have to satisfy the $MS-S-1$ relations (\ref{matrixcond}) with $N=MS$. 
This is the main scheme to arrive at a Lagrange form of a $M\times M$ matrix polynomial $D(\zeta)$ of degree $S$ for given $MS>S+1$ roots $\zeta_1,\dots,\zeta _{MS}$ of det$D(\zeta)$ and  corresponding given eigenvectors (see (\ref{vecv})) $z_1,\dots,z _{MS}$. On the technical side, the choice of the $M-1$ vectors $y^{(\alpha)}_j$, see (\ref{Dj}), which are orthogonal to $z_j$ for each given $j$, is more conveniently made while dealing with specific applications. 


\section{The MTM model} 

Here we compute, by the Lagrange representation method, the explicit expression of  of the  Darboux matrix associated with the Lax pair of the MTM model. For the sake of simplicity we produce the explicit expression of the lowest degree Darboux matrix polynomial which is compatible with the group of symmetries of this model. The construction of higher degree Darboux  matrices requires more computational efforts and it is not considered here. In laboratory coordinates $x$ (space) and $t$ (time) the MTM equations are
\begin{equation}\label{mtmlab}
i(u_t-cu_x) +\mu v=\frac{1}{\mu} |v|^2 u\;\;,\;\; i(v_t+cv_x) +\mu u=\frac{1}{\mu} |u|^2 v\;\;.
\end{equation}
Here $u=u(x,t)$ and $v=v(x,t)$ are the complex dependent variables, while $\mu$ is a constant real parameter and $c$ is a constant characteristic velocity. Although, by rescaling $x,t,u,v$, one could set $c=\mu=1$ we prefer to keep theses parameters as arbitrary. As for the integrability and relativistic spinor formulation of (\ref{mtmlab}), see \cite{M, KM}. We note that in relativistic field theory the parameter $\mu$ plays the role of the mass while in optics is related to the medium periodic (f.i. Bragg grating) constant \cite{MAKKC}. We also note that, like for the Nonlinear Schr\"{o}dinger equation to which the MTM equations (\ref{mtmlab}) reduce in the non relativistic limit, the nonlinearity is cubic and describes only cross-interaction (the addition of a cubic self-interaction term destroys both integrability and relativistic invariance). 

\noindent
 In order to write down the Lax pair,  we find it more convenient to use the light-cone coordinates $\xi=(ct+x)/(2c)\,,\,\eta=(ct-x)/(2c)$. In these coordinates the MTM equations (\ref{mtmlab}), which become 
\begin{equation}\label{mtmlight}
iu_\eta +\mu v=\frac{1}{\mu} |v|^2 u\;\;,\;\; iv_\xi +\mu u=\frac{1}{\mu} |u|^2 v\;\;,
\end{equation}
follow from the condition that the two ODEs 
 \begin{equation}\label{lax}
\Psi_\xi=A(\zeta) \Psi\;\;,\;\; \Psi_\eta=B(\zeta) \Psi\;\;,
\end{equation}
 be compatible with each other. The pair of matrices $A(\zeta)$ and $B(\zeta)$, as well as the solution $\Psi$ of the Lax equations (\ref{lax}), depend on the spectral complex parameter $\zeta$ and on the coordinates $\xi,\eta$. All these matrices are $2\times 2$, and $A(\zeta)$ and $B(\zeta)$ take the rational expression
 \begin{equation}\label{ABpair}
A(\zeta) =\frac{i\mu}{2} \zeta^2 \sigma_3+ \zeta U + \frac{i}{2\mu} |u|^2 \sigma_3 \;,\; B(\zeta) =\frac{i\mu}{2} \zeta^{-2} \sigma_3+ \zeta^{-1} V + \frac{i}{2\mu} |v|^2 \sigma_3\;, \;\;\;
\end{equation}
where $\sigma_3$ is one of the Pauli matrices
\begin{equation}\label{pauli}
\sigma_1=\left(\begin{array}{cc} 0 & 1 \\ 1 & 0 \end{array}\right ) \;,\; \sigma_2=\left(\begin{array}{cc} 0 & -i \\ i & 0\end{array} \right ) \;,\;  \sigma_3=\left(\begin{array}{cc} 1 & 0 \\ 0 & -1\end{array} \right ) \;.
\end{equation}
In these expressions the matrices $U=U(\xi,\eta)$ and $V=V(\xi,\eta)$ are Hermitian and off-diagonal,
\begin{equation}\label{UV}
U=\left(\begin{array}{cc} 0 & u^* \\ u & 0 \end{array}\right ) \;,\; V=\left(\begin{array}{cc} 0 & v^* \\ v & 0 \end{array}\right ) \;,
\end{equation}
whose entries are a solution of the MTM equations (\ref{mtmlight}) (an asterisk stands for complex conjugation). Because of the very special expressions (\ref{ABpair}), it is clear that  the MTM equations (\ref{mtmlight}) are a reduction of a larger system of equations \cite{DHS,BGK} containing more field functions rather than just two, i.e. $u,v$. Indeed the reduction conditions  can be obtained via the reduction method  based on automorphic matrix valued functions \cite{LM} as applied to the dihedral group \cal{D}$_2$. This can be easily realized by looking first at the  representation of  \cal{D}$_2$ whose four elements are generated by two reflections on the complex plane, namely
\begin{equation}\label{planed2}
g_0(\zeta)=\zeta\;\;,\;\;g_1(\zeta)=\zeta^*\;\;,\;\;g_2(\zeta)=-\zeta^*\;\;,\;\;g_3(\zeta)=-\zeta\;\;,
\end{equation}
where $\zeta$ is any non vanishing strictly complex number so that the rectangle $\{\zeta,\zeta^*,-\zeta^*,-\zeta \}$ does not degenerate into a segment. An automorphic function $f(\zeta)$ on the complex plane is a meromorphic function which satisfies the symmetry conditions on the group orbit (alias the rectangle (\ref{planed2}))
\begin{equation}\label{functd2}
G_j \cdot f(g_j(\zeta))=f(\zeta)\;\;,\;\;j=0,1,2,3 \;\;,
\end{equation}
where the four operators $G_j$, which are a representation of \cal{D}$_2$, are chosen to our purpose as $G_0\cdot z=G_3 \cdot z=z\,,\,G_1\cdot z=G_2 \cdot z=z^*$. Next we consider an automorphic matrix valued function $F(\zeta)$ as meromorphic and satisfying the symmetry relations $S_j[F(\zeta)]=F(\zeta)\,,\,j=0,1,2,3$. Here the four operators $S_j$ are the following representation of \cal{D}$_2$
\begin{equation}\label{matrixd2}
S_j[F(\zeta)]=\sigma_j G_j \cdot F(g_j(\zeta)) \sigma_j^{-1}\;\;,\;\;j=0,1,2,3 \;\;,
\end{equation}
where, in addition to the Pauli matrices (\ref{pauli}), we have introduced the  unit matrix $\sigma_0=1$. Since both matrices $A(\zeta), B(\zeta)$ in the Lax equations (\ref{lax}) are automorphic, 
 \begin{equation}\label{ABauto}
S_j[A(\zeta)]=A(\zeta)\;,\;S_j[B(\zeta)]=B(\zeta)\;\;,\;\;j=0,1,2,3 \;\;,
\end{equation}
the following observations turn out to be useful:

\noindent
\emph{Remark 1} : if the matrix $F(\zeta)$ depends on a variable $y$, then
\begin{equation}\label{deriv}
S_j[F_y(\zeta)]=\left(S_j[F(\zeta)]\right)_y\;\;,\;\;j=0,1,2,3 \;\;.
\end{equation} 
\emph{Remark 2} : for any two matrices $F_1(\zeta)$ and $F_2(\zeta)$
\begin{equation}\label{mult}
S_j[F_1(\zeta)\, F_2(\zeta)]= S_j[F_1(\zeta)] \,S_j[F_2(\zeta)] \;\;,\;\;j=0,1,2,3 \;\;.
\end{equation} 
\emph{Remark 3} : if $\Psi(\zeta)$ is a fundamental matrix solution of the Lax equations (\ref{lax}), then 
\begin{equation}\label{laxsymm}
S_j[\Psi(\zeta)]=\Psi(\zeta) \Gamma_j(\zeta) \;,\;j=0,1,2,3 \;\;,\;\Gamma_{j\xi}=\Gamma_{j\eta}=0\;,\;\textrm{det}\Gamma_j\neq 0\;\;.
\end{equation}
\emph{Remark 4} : by introducing the additional operator
\begin{equation}\label{Soper}
S[F(\zeta)]=-\sigma_3 F^\dagger(\zeta^*) \sigma_3 \;\;,
\end{equation}
and by noticing that its action on traceless matrices (Tr$F=0$) coincides with that of $S_1$, namely $S[F(\zeta)]=S_1[F(\zeta)]$, one concludes that the solution $\Psi(\zeta)$ of the Lax equations (\ref{lax}) is such that the matrix $\sigma_3 \Psi^\dagger(\zeta^*) \sigma_3 \Psi(\zeta)$ is $\xi$- and $\eta$-independent. 
 
While remarks 1 and 2 are strait consequences of the definition (\ref{matrixd2}), remark 3 follows from remarks 1 and 2 as applied to the Lax equations (\ref{lax}) together with the property $S_j \cdot S_j =1$ which implies that, if $\Psi(\zeta)$ is fundamental (say det$\Psi(\zeta)\neq 0$) then also $S_j[\Psi(\zeta)]$ is fundamental. The conclusion of remark 4 follows from the fact that the two matrices $A(\zeta)\,,\,B(\zeta)$ are traceless (see (\ref{ABpair})). 

\noindent
Let us consider now the DT $\Psi^{(1)}\rightarrow \Psi^{(2)}$ characterized by the Darboux matrix $D(\zeta)$ as follows
 \begin{equation}\label{Dmatrix}
\Psi^{(2)}(\zeta)=D(\zeta) \Psi^{(1)}(\zeta) \;\;,
\end{equation}
with the requirement that the Lax pair (\ref{lax}) transforms itself in covariant manner. This means that $\Psi^{(l)}\,,\,l=1,2$, is a matrix solution of (\ref{lax}) with $A(\zeta)=A^{(l)}(\zeta)$ and $B(\zeta)=B^{(l)}(\zeta)$ where $A^{(l)}(\zeta)$ and $B^{(l)}(\zeta)$ have the same expression (\ref{ABpair}) with $u,v$ replaced by $u^{(l)},v^{(l)}$. As a consequence $(u^{(1)},v^{(1)})$ and $(u^{(2)},v^{(2)})$ are two different solutions of the MTM equations (\ref{mtmlight}), and their relation with each other is induced by the DT (\ref{Dmatrix}) itself (see below). On the other hand the Darboux matrix $D(\zeta)$ depends on the coordinates $\xi,\eta$ according to the two compatible differential equations
\begin{equation}\label{Dlax}
D_\xi + D A^{(1)}- A^{(2)} D=0 \;\; \;,\;\;\;
D_\eta + D B^{(1)}- B^{(2)} D=0  \;\;.
\end{equation}
For future reference, we note here that, if $\hat{D}(\zeta)$ is a particular solution of these ODEs (\ref{Dlax}), then the general solution has the expression $D(\zeta)=\hat{D}(\zeta) \Psi^{(1)} \Gamma(\zeta) \Psi^{(1)-1}$ where $ \Gamma(\zeta)$ is an arbitrary constant (i.e. $\xi$- and $\eta$-independent) matrix. Again we observe, in analogy with remark 3 above, that

\noindent
 \emph{Remark 5} : if $D(\zeta)$ is a solution of the pair of equations (\ref{Dlax}) then 
 \begin{equation}\label{Dlaxsymm}
S_j[D(\zeta)]=D(\zeta) \Psi^{(1)} \Gamma_j(\zeta) \Psi^{(1)-1} \;,\;j=0,1,2,3 \;\;,\;\Gamma_{j\xi}=\Gamma_{j\eta}=0\;,\;\textrm{det}\Gamma_j\neq 0\;\;.
\end{equation}
This follows from the same arguments as for remark 3. We now assume that the Darboux matrix is polynomial in $\zeta$, and observe that

\noindent
\emph{Remark 6} : if $D(\zeta)$ has a polynomial dependence on $\zeta$ then also $S_j[D(\zeta)]$ does with the same degree, as it follows from the very definition (\ref{matrixd2}).

\noindent
\emph{Remark 7} : if the Darboux matrix $D(\zeta)$ is polynomial, then  $D(\zeta)$ is automorphic, namely
 \begin{equation}\label{Dauto}
S_j[D(\zeta)]=D(\zeta)  \;,\;j=0,1,2,3 \;\;.
\end{equation}
This is implicit in the fact that $ \Psi^{(1)} \Gamma_j(\zeta) \Psi^{(1)-1}$ (see (\ref{Dlaxsymm})) can only be a $\zeta$-independent scalar constant and that the polynomial Darboux matrix $D(\zeta)$ is defined modulo a scalar  factor. In this respect we note that this factor may be $j$-dependent and be a representation of the dihedral group. However  the effect of this factor turns out to be irrelevant.  Finally, in analogy with remark 4, we conclude with 

\noindent
\emph{Remark 8} :
\begin{equation}\label{Dproperty}
\sigma_3 D^\dagger(\zeta^*)\sigma_3 D(\zeta)= P(\zeta) \; \;,
\end{equation}
where $P(\zeta)$ is the (scalar) automorphic polynomial
\begin{equation}\label{Ppoly}
  P(\zeta)= \textrm{det}D(\zeta) \; \;.
\end{equation}
It is plain from (\ref{functd2})  that, if $\chi$ is a root of an automorphic polynomial, then all four points $g_j(\chi)$ of its associated \cal{D}$_2$ orbit are roots. Since we assume here and in the following that all roots are strictly complex (say no rectangle $g_j(\chi)$ is degenerate) and simple, the roots of the polynomial (\ref{Ppoly}) come in quadruplets and therefore the polynomial (\ref{Ppoly}) has the following representation
\begin{equation}\label{Proot}
  P(\zeta)=\prod_{n=1}^L (\zeta-\chi_n)(\zeta-\chi_n^*)(\zeta+\chi_n^*)(\zeta+\chi_n)  \; \;.
\end{equation}
Thus the polynomial (\ref{Ppoly}) has degree $4L$ since its roots are the vertices of $L$ rectangles. Moreover the Darboux matrix $D(\zeta)$, which is $2\times 2$, has degree $2L$,
\begin{equation}\label{DpolyL}
  D(\zeta)=\sum_{k=0}^{2L} D^{(k)} \zeta^k  \; \;.
\end{equation}
It is worth pointing out that the monodic form (\ref{Proot}) of the polynomial $P(\zeta)$ follows from the commutation relation $[\sigma_3\,,\,D^{(2L)}]=0$, see below, and from (\ref{Dproperty}).

\noindent
 At each root $g_j(\chi_n)$ of 
$P(\zeta)$ the matrix $D_j^{(n)}=D(g_j(\chi_n))$ has the eigenvector $z_j^{(n)}$ corresponding to the vanishing eigenvalue (see (\ref{Ppoly}))
\begin{equation}\label{eigenv}
   D_j^{(n)} z_j^{(n)} =0\; \;,\; j=0,1,2,3\;,\;n=1,\dots, L\;.
\end{equation}
Because of the symmetry relations (\ref{Dauto}), for each quadruplet $g_j(\chi_n)$ associated with the root $\chi_n$ the corresponding four matrices $D_j^{(n)}$
 are  related to each other. In fact, as a strait consequence of (\ref{Dauto}) for $\zeta= \chi_n$ together with the definition (\ref{matrixd2}), we obtain the following relations within each quadruplet (i.e. for fixed $n$)
\begin{equation}\label{symmatr}
   D_j^{(n)}  = \sigma_jG_j\cdot (D_0^{(n)}) \sigma_j\; \;,\; j=0,1,2,3\;\;,
\end{equation}
By assuming that the roots $\chi_n$ and the corresponding eigenvectors $z_0^{(n)}$ are given, we proceed to consider the Lagrange representation of the Darboux matrix $D(\zeta)$ according to the prescription given in the previous section 2. Moreover, in order to illustrate this construction of $D(\zeta)$ in the simplest possible way, we treat the case in which only one quadruplet is given, namely $L=1$ with the notation: $\chi_1=\chi$, $z_j^{(1)}=z_j$. Thus, since the rank $M$ and degree $S$ of the Darboux matrix $D(\zeta)$ is 2, the Lagrange form (\ref{lagrangeinter}), which in the present context reads
\begin{equation}\label{lagrangedarboux}
D(\zeta)= \sum_{j=1}^4 L^{(4)}_j(\zeta) D_{j-1}  \;\;,\;\;D_j=D(g_j(\chi))\;\;,
\end{equation} 
is a 4-point representation. Here the definition (\ref{lagrange}) applies with $\zeta_j = g_{j-1}(\chi)\,,\,j=1,\dots,4$ or, more explicitly, $\zeta_1=\chi\,,\, \zeta_2=\chi^*\,,\,\zeta_3=-\chi^*\,,\,\zeta_4=-\chi$. As pointed out in section 2, comparing the degree of the left-hand-side with that of the right-hand-side of this equation (\ref{lagrangedarboux}) yields just one condition, see (\ref{matrixcond}) with $S=2$ and $N=4$, on the matrices $D_j$, which is 
\begin{equation}\label{cond}
 \sum_{j=1}^4 C^{(4)}_{j4} D_{j-1} =0 \;\;.
\end{equation}  
Next we go to the expression (\ref{Dj}) of the four matrices $D_j$, which takes the dyadic form $D_j=w_j y_j^\dagger$. Here the vector $y_j$, which is orthogonal to  the eigenvector $z_j$ according to the prescription (\ref{orthog}), may be chosen as
\begin{equation}\label{yvec}
 y_j = \sigma_2 z_j^* \;\;.
\end{equation} 
Since we have to deal only with the matrix $D_0$ to perform our computing because of the symmetry relation (\ref{symmatr}), it remains to find the unknown vector $w_0$. To this aim, the starting point  is the condition (\ref{cond}), which explicitly reads
\begin{equation}\label{explicond}
 C^{(4)}_{14} D_{0}+ C^{(4)}_{24}\sigma_1 D_{0}^*\sigma_1+ C^{(4)}_{34} \sigma_2 D_{0}^*\sigma_2+ C^{(4)}_{44} \sigma_3 D_{0}\sigma_3=0 \;\;,
\end{equation}  
 together with the expressions $C^{(4)}_{14}=-C^{(4)}_{44}=[2\chi(\chi^2-\chi^{*2})]^{-1}\,,\,C^{(4)}_{24}=-C^{(4)}_{34}=C^{(4)*}_{14}$, and the final result is 
 \begin{equation}\label{w0}
   w_0  = \rho \chi  y_0 = \rho \chi  \sigma_2 z_0^* \;\;,
\end{equation}
where the normalizing factor $\rho$ is real and positive, but still to be found. The details of this computation, and of similar ones in the following, are omitted as lengthy and simple. Next, the Lagrange form (\ref{lagrangedarboux}) of $D(\zeta)$ and its power expansion (\ref{Dpoly}) (for $L=1$) imply the expressions (see also (\ref{matrixrel}) for $M=S=2$)
\begin{equation}\label{Dcoeff}
D^{(k)}  =\sum_{j=1}^4 C^{(4)}_{jk+1}\sigma_jG_j\cdot (D_0) \sigma_j\; \;,
\;\;k=0,1,2\;\;,
\end{equation}
where the Lagrange coefficients (see (\ref{lagrangecoeff})) take the following expressions $C^{(4)}_{13}=C^{(4)}_{43}=[2(\chi^2-\chi^{*2})]^{-1}\,,\,C^{(4)}_{23}=C^{(4)}_{33}=C^{(4)*}_{13}\;,\; C^{(4)}_{12}=-C^{(4)}_{42}=-\chi^{*2}[2\chi(\chi^2-\chi^{*2})]^{-1}\,,\,C^{(4)}_{22}=-C^{(4)}_{32}=C^{(4)*}_{12}\;,\; C^{(4)}_{11}=C^{(4)}_{41}=-\chi^{*2}[2(\chi^2-\chi^{*2})]^{-1}\,,\,C^{(4)}_{21}=C^{(4)}_{31}=C^{(4)*}_{11} $. The upshot of these computations is given by the following expressions
\begin{subequations}\label{expcoeff}
\begin{equation}\label{D2}
   D^{(2)}  = -\frac{\rho}{2(\chi^2-\chi^{*2})} \sigma_3 (\chi^*  \{ \sigma_3\,,\,z_0z_0^\dagger \}+\chi \sigma_1\{ \sigma_3\,,\,z_0z_0^\dagger \} \sigma_1 )\; \;,
\end{equation}
\begin{equation}\label{D1}
 D^{(1)}  = -\frac12 \rho \sigma_3  [ \sigma_3\,,\,z_0z_0^\dagger ] \;\;,
\end{equation}
\begin{equation}\label{D1}
 D^{(0)}  = |\chi|^2 D^{(2)\dagger} \;\;.
\end{equation}
\end{subequations}
Here $\{\bullet\,,\,\bullet\}$ and $[\bullet\,,\,\bullet]$ stand for the anticommutator and, respectively, for the commutator. We also note that, as it follows from remark $8$ (\ref {Dproperty}), the matrix coefficient $D^{(2)}$ turns out to be diagonal and unitary, $D^{(2)\dagger}\,D^{(2)}=1$, while the coefficient $D^{(1)}$ is off-diagonal and Hermitian, $D^{(1)\dagger}=D^{(1)}$. By a further computational effort, the matrix coefficients $D^{(2)}$ and $D^{(1)}$ can be given the explicit expressions
\begin{equation}\label{matrixD2}
   D^{(2)}  = \rho \frac{|\chi^* |z_{01}|^2 +\chi |z_{02}|^2|}{2 \textrm{Im}\chi^2}e^{i\phi\sigma_3} \; , \; e^{i\phi}=i\frac {\chi^* |z_{01}|^2 +\chi |z_{02}|^2}{|\chi^* |z_{01}|^2 +\chi |z_{02}|^2|}\;,
 z_0=\left ( \begin{array}{c} z_{01} \\ z_{02} \end{array} \right ) \;\;,
\end{equation}
\begin{equation}\label{matrixD1}
   D^{(1)}  = \left ( \begin{array}{cc} 0 & \delta^* \\ \delta & 0 \end {array} \right ) \; , \; \delta = -\rho z_{01}^* z_{02} \;\;,
\end{equation}
where the function $\phi=\phi(\xi,\eta)$ is real. We are now in the position to compute the value of the normalizing factor $\rho$, see (\ref{w0}). Indeed from the unitarity property of $D^{(2)}$ it follows that
\begin{equation}\label{ro}
 \rho=  \frac{2 \textrm{Im}\chi^2}{|\chi^* |z_{01}|^2 +\chi |z_{02}|^2|}  \; \;.
\end{equation}
Here we have conveniently, and with no loss of generality, set $\chi$ in the first quadrant of the complex plane, say $\textrm{Re}\chi >0$ and  $\textrm{Im}\chi >0$.

\noindent 
The DT transformation (\ref{Dmatrix}), which acts on the solution $\Psi^{(1)}$ of the Lax pair, obviously induces the transformation $(u^{(1)}\,,\,v^{(1)})\rightarrow (u^{(2)}\,,\,v^{(2)})$ on the corresponding solutions of the MTM equations. This transformation is derived from the ODEs (\ref{Dlax}) for the Darboux matrix itself $D(\xi,\eta,\zeta)$ by looking at the coefficient of the power $\zeta^3$ in the first equation and at the coefficient of the power $\zeta^{-1}$ in the second equation. This simple derivation leads to the following matrix transformation
\begin{equation}\label{matrixBT}
U^{(2)}  =D^{(2)} U^{(1)} D^{(2)\dagger} -i\mu \sigma_3 D^{(1)}D^{(2)\dagger}  \; \;,
\;\;V^{(2)}  =D^{(2)\dagger} V^{(1)} D^{(2)} -\frac{i\mu}{|\chi|^2} \sigma_3 D^{(1)}D^{(2)} \;\;,
\end{equation}  
or, more explicitly (see (\ref{UV})),
\begin{equation}\label{BT}
 u^{(2)}=  e^{-i\phi} ( e^{-i\phi} u^{(1)} +i\mu \delta) \; \;,\;\;  v^{(2)}=  e^{i\phi} ( e^{i\phi} v^{(1)} +i\frac{\mu}{|\chi|^2} \delta) \;\;.
 \end{equation}
This  concludes the algebraic side of the construction of the Darboux transformation. At last we turn our attention to the differential part by deriving the  
$\xi$- and $\eta$-dependence of the vector $z_0$. To this purpose we consider the ODEs (\ref{Dlax}) for $\zeta=\chi$, each term being applied to the eigenvector $z_0$, see (\ref{eigenv}) with $j=0\,,\,n=1$. As a consequence $D_0(-z_{0\xi} + A^{(1)}(\chi) z_0)=0$ and $D_0(-z_{0\eta} + B^{(1)}(\chi) z_0)=0$, and therefore, without any loss of generality, the vector $z_0$ can be characterized as a solution of the Lax pair corresponding to the spectral parameter $\zeta=\chi$ and to the solution $(u^{(1)}\,,\,v^{(1)})$ of the MTM. Thus its expression is known,
\begin{equation}\label{v0}
z_0 = \Psi^{(1)}(\xi,\eta, \chi) \gamma  \;\;,
 \end{equation}
 where only the constant vector $\gamma$ is left arbitrary. Indeed the solution $(u^{(1)}(\eta, \chi)\,,\,v^{(1)}(\eta, \chi))$, together with its corresponding matrix $\Psi^{(1)}(\zeta,\eta, \chi)$, are considered as known. 


\section{Conclusions} 
 Darboux transformations have been recognized as a useful tool to obtain the  expression of solutions describing soliton propagation over an arbitrary known background. In this spirit we have presented a novel algorithmic way to construct polynomial Darboux matrices in Lagrange representation (\ref{lagrangeinter}) rather than in standard power form (\ref{poly}). A good side of our technique is that it requires computations in the algebra of matrices with no need to deal explicitly with specific matrix entries and large linear systems. As a difference, and advantage, with respect to other approaches, we note that also the matrix coefficient of the highest power of the Darboux matrix $D(\zeta)$ is found by purely algebraic computation. This is in contrast with the assumption that this coefficient is often set to be unit or, whenever this assumption cannot be valid, it has to be evaluated by integrating an ODE, as it happens in alternative methods. Moreover our approach is particularly suited to take into account the symmetries of the associated Lax pair. We deem all the main features of our way of constructing Darboux matrices are well illustrated by the MTM study case of section 3. The technical problem which has not been treated here is the extension of our method to  construct N-fold Darboux transformations.

\end{document}